\documentclass[aps,prd,twocolumn,showpacs,nofootinbib]{revtex4}

\usepackage{graphicx,amssymb,bm}

\begin{document}

\title{Anisotropy of the cosmic gamma-ray background from dark matter
annihilation}

\author{Shin'ichiro Ando}
\email{ando@utap.phys.s.u-tokyo.ac.jp}
\affiliation{Department of Physics, School of Science, University of
Tokyo, Tokyo 113-0033, Japan}

\author{Eiichiro Komatsu}
\email{komatsu@astro.as.utexas.edu}
\affiliation{Department of Astronomy, University of Texas at Austin,
Austin, TX 78712}

\date{December 8, 2005; accepted January 13, 2006}

\begin{abstract}
High-energy photons from pair annihilation of dark matter particles
 contribute to the cosmic gamma-ray background (CGB) observed in a wide
 energy range.
Since dark matter particles are weakly interacting, annihilation can
 happen only in high density regions such as dark matter halos.
The precise shape of the energy spectrum of CGB depends on the nature of
 dark matter particles --- their mass and annihilation cross section, as well
 as the cosmological evolution of dark matter halos.
In order to discriminate between the signals from dark matter
 annihilation and other astrophysical sources, however, the information
 from the energy spectrum of CGB may not be sufficient.
We show that dark matter annihilation not only contributes to the mean
 CGB intensity, but also produces a characteristic {\it anisotropy},
 which provides a powerful tool for testing the origins of the observed
 CGB.
We develop the formalism based on a halo model approach to analytically
 calculate the three-dimensional power spectrum of dark matter clumping,
 which determines the power spectrum of annihilation signals.
We show that the expected sensitivity of future gamma-ray detectors such
 as the Gamma Ray Large Area Space Telescope (GLAST) should allow us to
 measure the angular power spectrum of CGB anisotropy, if dark matter
 particles are supersymmetric neutralinos and they account for most of
 the observed mean intensity of CGB in GeV region.
On the other hand, if dark matter has a relatively small mass, on the
 order of 20 MeV, and accounts for most of the CGB in MeV region, then
 the future Advanced Compton Telescope (ACT) should be able to measure
 the angular power spectrum in MeV region.
As the intensity of photons from annihilation is proportional to the
 density squared, we show that the predicted shape of the angular power
 spectrum of gamma rays from dark matter annihilation is different
 from that due to other astrophysical sources such as blazars and
 supernovae, whose intensity is linearly proportional to density.
Therefore, the angular power spectrum of the CGB provides a
 ``smoking-gun'' signature of gamma rays from dark matter annihilation.
While the mean CGB intensity expected from dark matter halos with smooth
 density profiles is smaller than observed, the dark matter substructure
 within halos may provide the origin of additional ``boost'' factors for
 the annihilation signal.
Our formalism can be used for any other radiation processes that involve
 collision of particles.
\end{abstract}

\pacs{%
95.35.+d;   
95.85.Pw;   
98.70.Vc    
}

\maketitle


\section{Introduction}
\label{sec:Introduction}

The energy density of the universe is dominated by invisible components:
dark matter and dark energy, both of which are of unknown origin.
Remarkable progress in both the theoretical and observational studies of
the cosmic microwave background (CMB) anisotropy, Type Ia supernovae,
and large-scale structure has allowed us to precisely determine what
fraction of the total energy these dark components convey, $\Omega_\chi
= 0.23$ and $\Omega_\Lambda = 0.73$ \cite{WMAP}, where the subscripts
$\chi$ and $\Lambda$ denote dark matter particles and dark energy,
respectively.

While the nature of dark energy is a complete mystery, there are several
candidates for dark matter particles.
Of which, the most popular candidate is the stable supersymmetric
neutralino with mass on the order of GeV to TeV, which can explain the
observed dark matter mass density today.
Although very weak, it is expected that dark matter particles may
interact with the usual matter via scattering, interact with themselves,
and/or annihilate into gamma rays, positrons, and neutrinos.
The direct detection of particle dark matter is, therefore, under
intensive and extensive efforts of both physicists and astronomers
\cite{DMReview1,DMReview2,DMReview3}.

High-energy photons from annihilation of dark matter particles provide
indirect means to probe the properties of dark matter.
Annihilation signatures, especially gamma rays, have been searched for
in regions where the dark matter density is expected to be high, as
annihilation rate is proportional to the density squared,
$\rho_\chi^2$.
An obvious site is the central region of our Galaxy.
Strong gamma-ray emission has been detected toward the Galactic center
over a wide energy range, and the nature and origin of this emission
have been investigated by many researchers 
\cite{DMGammaGC1,DMGammaGC2,DMGammaGC3,DMGammaGC4,DMGammaGC5,DMGammaGC6,DMGammaGC7,DMGammaGC8,DMGammaGC9}.
Although both the spectrum and angular distribution of gamma rays from
the Galactic center are well observed, it is still quite difficult to
distinguish a dark matter component from the other possibilities such as
emission from ``ordinary'' astrophysical objects; even future gamma-ray
detectors and telescopes do not have sufficient resolution to remove all
the other sources from the Galactic central image.

Another possibility is the extragalactic background light, the cosmic
gamma-ray background (CGB), which has been measured in a wide energy
range \cite{GeVCGB1,GeVCGB2,MeVCGB1,MeVCGB2,MeVCGB3}.
It has been speculated that some fraction of the CGB may originate from
annihilation of dark matter particles in halos distributed over
cosmological distances
\cite{CGBDM1,CGBDM2,CGBDM3,CGBDM4,CGBDM5,A05,OTN05,AK1,AK2}.
The dark matter contribution to the CGB depends on the nature of dark
matter particles, such as their mass and annihilation cross section, as
well as the cosmological evolution of dark matter halos.
While it is likely that the dominant part of the CGB spectrum in the GeV
region, detected by the Energetic Gamma Ray Experiment Telescope (EGRET)
\cite{GeVCGB1,GeVCGB2}, comes from unresolved blazars, i.e., beamed
population of Active Galactic Nuclei (AGN), annihilating dark matter
might give large contribution at some specific energy
range.\footnote{As has been known and is shown below, with the canonical
choice of relevant parameters, dark matter annihilation cannot give
enough contribution to the mean CGB intensity; one still needs large
boost by some mechanism. In addition, its contribution is strongly
constrained by gamma-ray observations toward the
Galactic center \cite{A05}. However, one might avoid these constraints by 
invoking gamma-ray emission from sub-halos having 
$M\sim 10^{-6} M_\odot$ within host halos \cite{OTN05},
or from density cusps forming around
intermediate-mass black holes \cite{IMBH}.}
Possible candidates for this case include the supersymmetric neutralino,
as well as the Kaluza-Klein dark matter predicted by theories of universal
extra-dimension (e.g., \cite{DMReview3}) and heavy relic neutrinos
\cite{Neutrino1,Neutrino2}.
On the other hand, the origin of the CGB in MeV region is much less
understood than in GeV region.
The soft gamma-ray spectrum in 1--20 MeV cannot fully be attributed to
either AGN or Type Ia supernovae or a combination of the two
\cite{SNIa1,SNIa2}; thus, annihilation of dark matter particles is one
of the most viable explanations for the CGB in MeV region, provided that
the dark matter mass is around 20 MeV \cite{AK1,AK2}.
Such a light dark matter particle was originally introduced to explain
the origin of the 511 keV emission line from the Galactic center
\cite{MeVDM,BBB}, detected by the International Gamma-Ray Astrophysics
Laboratory (INTEGRAL) \cite{INTEGRAL1,INTEGRAL2}, which is otherwise
difficult to explain.
Therefore, the MeV dark matter is an attractive candidate that satisfies
the observational constraints from both the Galactic center and the
CGB,\footnote{After this paper has been submitted, 
Ref.~\cite{beacom/yuksel:2005} has appeared on the preprint server.
They claim that the mass of MeV dark matter should be less than
3 MeV from a more accurate treatment of relativistic positrons
annihilating with electrons in the interstellar medium.
While we use $m_\chi=20$~MeV throughout this paper, 
one can easily extend our calculations to arbitrary dark matter masses.}
while its particle physics motivation is less clear than for
the neutralinos.

Dark matter annihilation may be a viable explanation for the CGB, but how
do we know for sure that the CGB does come from annihilation?
What would be a smoking-gun signature for the annihilation signal?
We argue that {\it anisotropy} of the CGB may provide a smoking-gun
signature.
Although the CGB is isotropic at the leading order, anisotropy should
also exist if the CGB originates from cosmological halos.
The future gamma-ray detectors with an enhanced sensitivity and angular
resolution, such as the Gamma Ray Large Area Space Telescope (GLAST) or
the Advanced Compton Telescope (ACT), should be able to see such
anisotropy.
We believe that the CGB anisotropy is going to be the key to
discriminating between the dark matter annihilation signal and the other
sources.

In this paper, we develop the formalism to analytically calculate the
angular power spectrum of the CGB from dark matter annihilation.
We calculate the angular power spectrum in GeV region (for
supersymmetric neutralinos) as well as in MeV region (for MeV dark
matter).
We then discuss the detectability of CGB anisotropy in GeV region by
GLAST and in MeV region by ACT, showing that the predicted anisotropy
can be easily measured by 1-year operation of these experiments.
The formalism given in this paper can also be used to evaluate the CGB
anisotropy due to the other astrophysical objects with some
modification.
Note that the case of Type Ia supernovae has been discussed in
Ref.~\cite{SNIaAnisotropy}, being complementary to the present study.

This paper is organized as follows.
In Sec.~\ref{sec:Cosmic gamma-ray background: Isotropic component}, we
review the mean intensity of the CGB from dark matter annihilation.
In Sec.~\ref{sec:Cosmic gamma-ray background anisotropy}, we develop the
formalism to calculate the power spectrum of CGB anisotropy in the
context of the halo models. Results of the angular power spectrum are
then shown in Sec.~\ref{sec:Angular power spectrum} for both the
neutralino and MeV dark matter.
We compare the predictions with the expected sensitivities of the future
gamma-ray detectors.
In Sec.~\ref{sec:Discussion}, we discuss energy dependence of the
obtained anisotropy (Sec.~\ref{sub:Dependence on gamma-ray energy}),
other astrophysical sources which might contaminate the cosmological CGB
(Sec.~\ref{sub:Other astrophysical sources}), and the effects of dark
matter substructures (Sec.~\ref{sub:Substructure of dark matter
halos}).
In Sec.~\ref{sec:Conclusions}, we conclude the present paper with a
brief summary.


\section{Cosmic gamma-ray background: Isotropic component}
\label{sec:Cosmic gamma-ray background: Isotropic component}

Since the gamma-ray emissivity from dark matter annihilation is
proportional to the density squared, $\rho_\chi^2$, the CGB intensity
(the number of photons per unit area, time, solid angle, and energy
range) toward the direction $\hat{\bm n}$ can be generally expressed as
\begin{equation}
 I_\gamma (\hat{\bm n},E_\gamma) = \int dr~ \delta^2(r,\hat{\bm n}r)
  W([1+z]E_\gamma,r),
  \label{eq:intensity}
\end{equation}
where $E_\gamma$ is the observed gamma-ray energy, $z$ is the redshift,
$r$ is the comoving distance out to an object at $z$, 
$W$ is some function of gamma-ray energy and $r$ that is given below,
and $\delta$ is the overdensity at $\hat{\bm n}r$ compared to the
average dark matter mass density, $\bar{\rho}_\chi$, given by
\begin{equation}
 \delta\equiv \frac{\rho_\chi}{\bar{\rho}_\chi}.
\end{equation}
Because the dominant contribution comes from the dark matter halos,
$\delta$ is always much larger than unity.
Note that we label time by $r$ (or redshift $z$ used interchangeably),
and space by $\bm r = \hat{\bm n}r$.

We first derive the specific form of the function $W$.
The general formula for the intensity is given by \cite{Peacock}
\begin{eqnarray}
 E_\gamma I_\gamma (\hat{\bm n},E_\gamma) & = & \frac{c}{4\pi}\int dz~
  \frac{P_\gamma([1+z] E_\gamma,z,\hat{\bm n}r)}{H(z) (1+z)^4}
  \nonumber\\&&{}\times
  e^{-\tau ([1+z] E_\gamma,z)},
  \label{eq:general expression of intensity}
\end{eqnarray}
where $P_\gamma$ is the volume emissivity (energy of photons per unit
volume, time, and energy range), $H^2(z) \equiv H_0^2[\Omega_m (1+z)^3 +
\Omega_\Lambda]$ is the Hubble function in a flat universe, and we
assume the standard values for cosmological parameters, $H_0 =
100~h$ km s$^{-1}$ Mpc$^{-1}$ with $h = 0.71$, $\Omega_m = 0.27$, and
$\Omega_\Lambda = 0.73$.
By the exponential factor, we incorporate the effect of the gamma-ray
absorption due to pair creation with the diffuse extragalactic
background light in the infrared or optical bands \cite{Absorption};
however, this effect is actually negligible for the cases we consider
here where the gamma-ray energy is smaller than 1 TeV.
The annihilation rate per unit volume is given by $(\rho_\chi/m_\chi)^2
\langle \sigma v\rangle/2$, where $m_\chi$ is the mass of dark matter
particle and $\langle\sigma v\rangle$ is the annihilation cross section
times relative velocity averaged with the weight of the velocity
distribution.
The volume emissivity is therefore given by
\begin{equation}
 P_\gamma (E_\gamma,z,\hat{\bm n}r) = 
  E_\gamma\frac{dN_\gamma}{dE_\gamma}
  \frac{\langle \sigma v\rangle}{2}
  \left[\frac{\rho_\chi(z,\hat{\bm n}r)}{m_\chi}\right]^2,
  \label{eq:emissivity}
\end{equation}
where $dN_\gamma/dE_\gamma$ is the gamma-ray spectrum per annihilation.
Comparing Eq.~(\ref{eq:intensity}) with Eqs.~(\ref{eq:general expression
of intensity}) and (\ref{eq:emissivity}), and recalling that $dr = -c
dz/ H(z)$ and $\rho_\chi(z,\hat{\bm n}r) = \bar\rho_\chi(z)
\delta(z,\hat{\bm n}r) = \Omega_\chi \rho_c (1+z)^3\delta(z,\hat{\bm
n}r)$, we obtain $W$ as
\begin{equation}
 W(E_\gamma,z) = \frac{\langle\sigma v\rangle}{8\pi}
  \left(\frac{\Omega_\chi \rho_c}{m_\chi}\right)^2(1+z)^3
  \frac{dN_\gamma}{dE_\gamma} e^{-\tau(E_\gamma,z)}.
  \label{eq:W function}
\end{equation}

By taking the ensemble average of Eq.~(\ref{eq:intensity}), we obtain
the isotropic background component of the intensity, $\langle I_\gamma
(E_\gamma)\rangle$; it is equivalent to our evaluating $\langle\delta^2
(z)\rangle$, the mean clumping factor of dark matter density
fluctuations, which is given by the integral of two-point correlation
over the momentum.
We use a halo model approach (see Ref.~\cite{HaloModelReview} for a
review) to calculate a reduced part of the $N$-point correlation
function of density fluctuations. 
The clumping factor, $\langle\delta^2\rangle$, is then given
by\footnote{The expression~(\ref{eq:intensity multiplier}) is identical
to the 1-halo term $I_{11}^{(1)}$ of Eq.~(2.15a) in
Ref.~\cite{SB91}. The second (2-halo) term $I_{12}^{(2)}$ disappears,
because we evaluate the quantity at one point and halos are assumed to
be exclusive, having stiff boundary.}
\begin{eqnarray}
 \langle\delta^2(z) \rangle &=& \int_{M_{\rm min}}^\infty dM
  \frac{dn}{dM}(M,z)
  \left(\frac{M}{\Omega_m\rho_c}\right)^2
  \nonumber\\&&{}\times
  \int dV\  u^2 (r|M,z),
  \label{eq:intensity multiplier}
\end{eqnarray}
where $u(r|M,z) = \rho(r|M,z)/M$ is the density profile divided by the
halo mass $M$, $dn/dM$ the halo mass function for which we use the
expression given in Ref.~\cite{ST}.
This clumping factor $\langle \delta^2(z) \rangle$ is also known as an
intensity multiplier (it is the same quantity as, e.g., $\Delta^2(z)$ of
Ref.~\cite{CGBDM2} and $f(z)$ of Ref.~\cite{CGBDM3}).

We should comment on the boundary of the mass integral in
Eq.~(\ref{eq:intensity multiplier}).
As the mass function falls exponentially toward large masses, the
integral converges at high mass end; however, it actually diverges at
low mass end, although only very weakly.
This raises some questions regarding robustness of theoretical predictions
for the mean intensity of CGB from dark matter annihilation.
What determines the minimum mass, $M_{\rm min}$? 
In principle, one should integrate the mass function down to the
free-streaming scale or the Jeans mass of dark matter particles, below
which there is no fluctuation in dark matter. 
For both neutralinos and MeV dark matter, the minimum mass is roughly 
on the order of the mass of Earth \cite{Hofmann,Green1,Green2,Loeb}, 
while it can vary by orders of 
magnitude depending on the precise
value of the dark matter mass. (One should recalculate $M_{\rm min}$ as 
a function of the dark matter mass \cite{AK1} or any other 
particle physics parameters that affect the strength of interactions.)
When gamma rays emerge from annihilation without any other interactions
of the products of annihilation with the materials within halos 
--- i.e., dark matter annihilating directly into photons, 
photons emerge as the final state radiation of annihilation 
(also known as the internal bremsstrahlung), 
or decay of the products of annihilation into gamma rays
(e.g., decay of $\pi^0$ produced by annihilation of neutralinos), 
one should use the
minimum mass set by the free-streaming mass or the Jeans mass
of dark matter particles.
On the other hand, when gamma rays are produced predominantly 
by interactions between
the products of annihilation and the materials within halos after
annihilation, e.g., dark matter annihilation producing intermediate
particles such as positrons which, in turn, annihilate with electrons
{\it in the halo} to produce gamma rays, the minimum mass would be set by the
Jeans mass of {\it baryons}, which is on the order of $10^6M_\odot$
or larger \cite{rasera/etal:2005}.
Therefore, which minimum mass to use depends on what radiative processes
one assumes for the production of gamma rays per annihilation.

This issue is further complicated by the fact that 
not all dark matter halos would actually survive during the formation
of large-scale structure. Halos above the baryonic Jeans
mass would survive without any problems; however, 
a significant fraction of halos with very small free-streaming 
masses, say $10^{-6}M_\odot$, would have been tidally disrupted
shortly after their formation and might not be able to contribute
to the CGB. The critical mass above which the tidal disruption
due to hierarchical clustering becomes inefficient is difficult
to calculate accurately at 
present \cite{EarthDM,berezinsky/dokuchaev/eroshenko:2005}.
(Note that we are not concerned about the tidal disruption of 
these micro halos within larger halos --- this would be a subject
of the effect of substructures within halos, and we shall come back
to this point in Sec.~\ref{sub:Substructure of dark matter halos}.
Here, we are mostly concerned about the tidal disruption of
the extragalactic micro halos that are outside of larger halos.)

Generally, the dependence of CGB anisotropy on the minimum mass is
expected to be not very strong, when the amplitude of anisotropy 
is divided by the mean
intensity, $\delta I_\gamma / \langle I_\gamma \rangle$, the dependence
approximately canceling out in the numerator and denominator.
To take into account the uncertainty in our understanding of the 
minimum mass contributing to the CGB, we explore two 
cases, $M_{\rm min}=10^6 M_\odot$ and $10^{-6} M_\odot$, i.e.,
12 orders of magnitude variation in the minimum mass.

The clumping factor depends on $z$ through the halo mass function
\cite{ST} and the concentration of density profiles \cite{CONC1,CONC2}.
To compute the linear power spectrum that is necessary for the halo mass
function, we adopt the fitting formula in Ref.~\cite{LinearPS}.
The concentration parameter as a function of halo mass and redshift
determines how steeply dark matter distributes in a halo; we use the
result given in Ref.~\cite{CONC1}.
Using these ingredients, we compute the clumping factor for a given
density profile.
For the Navarro-Frenk-White (NFW) profile \cite{NFW1,NFW2} ($\gamma =
1$, where $\gamma$ is defined by $\rho \propto r^{-\gamma}$ for small
radii), we obtain $\langle\delta^2(0)\rangle \sim 4 \times 10^4$ for 
$M_{\rm min}=10^{6} M_\odot$ and $3 \times 10^5$ for $M_{\rm
min}=10^{-6} M_\odot$.
For steeper profiles such as the one with $\gamma = 1.5$ suggested by
the other simulations \cite{M99}, the clumping factor increases by
about an order of magnitude.
More recent simulations suggest that $\gamma$ depends on radii and could
be anywhere between 1 and 1.5 in the inner region
\cite{N-body1,N-body2,N-body3} (see also
Refs.~\cite{AnalyticProfile1,AnalyticProfile2,AnalyticProfile3} for
analytical approaches).
Again, while the mean intensity depends strongly on $\gamma$ (see, e.g.,
Ref.~\cite{AK1}), anisotropy (divided by the mean intensity) is expected
to be less sensitive to the precise value of $\gamma$.
For definiteness we adopt $\gamma=1$ throughout the paper.
Incidentally, it is practically useful to use the NFW profile, as many
quantities used in the formalism can be calculated analytically for this
profile.

One also needs to specify the particle physics parts involved in the
function $W$ --- the dark matter mass, averaged cross section $\langle
\sigma v \rangle$, and gamma-ray spectrum per annihilation.
We assume that the dark matter particle contributing to the CGB totally
accounts for the observed dark matter mass density, $\Omega_\chi =
0.23$.
In many cases, the annihilation cross section is closely related to
$\Omega_\chi$, as it determines the abundance of dark matter particles
having survived pair annihilation in the early universe.\footnote{This
would not be quite true when co-annihilations of dark matter particles
with other particles dominate (e.g., neutralinos co-annihilating with
other supersymmetric particles).
In such a case the annihilation cross section would have little 
to do with $\Omega_\chi$.
See Ref.~\cite{DMReview3} for more details.}
For both the supersymmetric neutralino and MeV dark matter, the
canonical value of the cross section is $\langle \sigma v \rangle =
3\times 10^{-26}$ cm$^3$ s$^{-1}$, while a wide range of parameter space
is still allowed\footnote{For MeV dark matter, it has been argued that
the canonical, velocity-independent cross section would overproduce
the Galactic 511 keV emission \cite{MeVDM,ascasiber}. However,
uncertainty in our understanding of the dark matter profile of our Galaxy 
still allows for the canonical cross section for 
$m_\chi\gtrsim 20$ MeV \cite{AK1,AK2}.
If, on the other hand, $m_\chi$ turns out to be less than 20~MeV, then
a smaller cross section may be preferred.};
in addition, we assume it to be independent of the
relative velocity.
Once again, the value of the annihilation cross section actually does
not affect the CGB anisotropy, when it is normalized by the mean
intensity, $\delta I_\gamma/\langle I_\gamma\rangle$.
(The cancellation is exact when it is independent of velocity.)
Thus, although a larger cross section is favored in order to make the
dark matter contribution to the CGB mean intensity more significant, the
prediction for the amplitude of anisotropy (divided by the mean
intensity) is robust regardless of the precise value of the cross
section.

While the energy spectrum of gamma-ray emission per annihilation
contains rich information about the nature of dark matter particles, we
do not pay much attention to the detailed shape of the spectrum, but
rather simply split the spectrum into two parts: MeV and GeV regions,
the former representing MeV dark matter, and the latter representing
neutralinos.
This is a reasonable approximation, as each contribution gives continuum
gamma-ray emission that is roughly constant over relevant energy
regions.
For neutralinos we use a simple parameterization, $dN_\gamma/dE_\gamma
\simeq (0.73/m_\chi) e^{-7.76E_\gamma/m_\chi} /
[(E_\gamma/m_\chi)^{1.5}+0.00014]$ \cite{CGBDM1}.
The Kaluza-Klein dark matter particle might also contribute to
the CGB in GeV region; their contribution may be calculated using
an appropriate gamma-ray energy spectrum per annihilation.
For MeV dark matter we consider the internal bremsstrahlung ($\chi \chi
\to e^+ e^- \gamma$) as the source of continuum emission, 
which gives $dN_\gamma / dE_\gamma
= \alpha [\ln (s^\prime / m_e^2) - 1] [1 + (s^\prime / s)^2] / (\pi
E_\gamma)$, where $\alpha = 1/137$ is the fine structure constant, $s =
4 m_\chi^2$, and $s^\prime = 4 m_\chi (m_\chi - E_\gamma)$ \cite{BBB}.
We shall ignore the 511 keV line emission, as its contribution 
must be subdominant compared with the AGN contribution \cite{AK1}.

\begin{figure}
\begin{center}
\includegraphics[width=8.5cm]{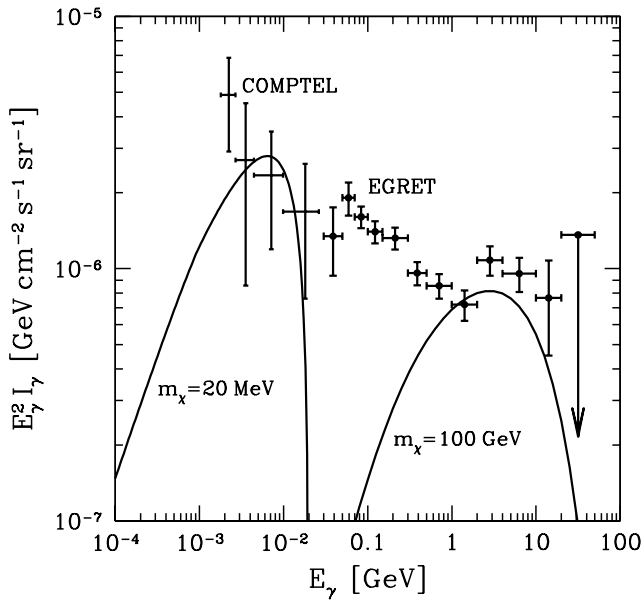}
\caption{Average intensity of the CGB from dark matter annihilation
 compared with the COMPTEL and
 EGRET data. The curve in GeV region is the predicted intensity
 for neutralinos with  $m_\chi = 100$ GeV, while that in MeV region 
 is the prediction for MeV dark matter particles with $m_\chi = 20$ MeV. As
 for the normalization, we have 
 used $\langle \sigma v \rangle = 3 \times 10^{-26}$
 cm$^3$ s$^{-1}$, and extra ``boost'' factors have been multiplied
 to match the observational data. For details of the boost factors,
 see the last paragraph of Sec.~II.}
\label{fig:CGB_spect}
\end{center}
\end{figure}

In Fig.~\ref{fig:CGB_spect}, we show the average CGB intensity
calculated with the ingredients given above for the neutralino ($m_\chi
= 100$ GeV) and MeV dark matter ($m_\chi = 20$ MeV).
We also show the observational data 
obtained by the Imaging Compton Telescope (COMPTEL)
\cite{MeVCGB3} and EGRET \cite{GeVCGB2}.
The predicted spectra have been multiplied 
by a factor of $1.5\times 10^3$ ($2.4\times 10^2$) for the neutralino,
and 10 (1.6) for the MeV dark matter, respectively,
assuming $M_{\rm min} = 10^6 M_\odot ~(10^{-6} M_\odot)$,
 to reasonably match the observational data.
These boost factors may originate from dark matter substructures within
halos or the cross section potentially larger than the value adopted
here.
Note that the 20 MeV dark matter with  
$M_{\rm min} =10^{-6} M_\odot$ (which is not very far away from 
the actual free-streaming mass of 20 MeV dark matter particles)
can already account for most of the observed CGB
without the need for a substantial boost due to
substructures (the boost factor is only 1.6), which agrees with
a more detailed analysis performed in Ref.~\cite{AK2}.

\section{Cosmic gamma-ray background anisotropy}
\label{sec:Cosmic gamma-ray background anisotropy}

In this section, we develop the formalism to analytically calculate the
angular power spectrum of CGB anisotropy due to annihilation of dark
matter particles.
We note that this formalism is quite general, and with some
modification, it can be applied to other situations including gamma rays
from ordinary astrophysical objects.

\subsection{General setup}
\label{sub:General setup}

We expand the deviation of CGB intensity from its mean value into the
spherical harmonics coefficients, $a_{lm}$,
\begin{equation}
 \delta I_\gamma (\hat{\bm n})
  \equiv I_\gamma (\hat{\bm n}) - \langle I_\gamma \rangle
  = \langle I_\gamma \rangle \sum_{lm} a_{lm} Y_{lm}(\hat{\bm n}).
  \label{eq:expansion with Y}
\end{equation}
Note that we have defined $a_{lm}$ as a dimensionless quantity; thus,
$a_{lm}$ represents the amplitude of anisotropy divided by the mean
intensity.
Using the orthonormal relation of $Y_{lm}(\hat{\bm n})$, one obtains
\begin{eqnarray}
 \langle I_\gamma \rangle a_{lm} &=& \int d \hat{\bm n} \ \delta
  I_\gamma (\hat{\bm n}) Y_{lm}^\ast (\hat{\bm n})
  \nonumber\\
 &=& \int d \hat{\bm n} \int dr\ f(r,\hat{\bm n}r) W(r)
  Y_{lm}^\ast (\hat{\bm n}),
  \label{eq:a_lm}
\end{eqnarray}
where
\begin{equation}
f \equiv \delta^2 - \langle \delta^2 \rangle,
\end{equation}
and the energy index in the function $W$ has been suppressed to simplify
the notation.

The goal of the present paper is to evaluate the angular power spectrum,
$C_l \equiv \langle |a_{lm}|^2 \rangle$, which is given by
\begin{equation}
 \langle I_\gamma \rangle^2 C_l = \int \frac{dr}{r^2}\
  \left\{W([1+z]E_\gamma,r)\right\}^2
  P_f \left( k = \frac{l}{r}; r\right).
  \label{eq:C_l}
\end{equation}
The detailed derivation of this result is given in
Appendix~\ref{eq:Relation between angular and three-dimensional power
spectrum}.
Here, $P_f(k)$ is the three-dimensional (3D) power spectrum of $f$,
defined by
\begin{equation}
  \langle \tilde f_{\bm k} \tilde f_{{\bm
k}^\prime} \rangle = (2\pi)^3 \delta^{(3)} (\bm k + \bm{k}^\prime)
P_f(k),
\end{equation}
where $\delta^{(N)}$ represents the $N$-dimensional delta function, and
$\tilde f_{\bm k}$ is the Fourier transform of $f(\bm r)$.
Note that $\bm k$ is a comoving wavenumber, as it is a Fourier variable
corresponding to the comoving distance $\bm r$.
Given this relation (\ref{eq:C_l}), we first focus on deriving the 3D
power spectrum, $P_f(k)$, and then turn to evaluating the angular power
spectrum.

\subsection{Three-dimensional power spectrum}
\label{sub:Three-dimensional power spectrum}

\subsubsection{Two-point contribution}

\begin{figure}
\begin{center}
\includegraphics[width=8.5cm]{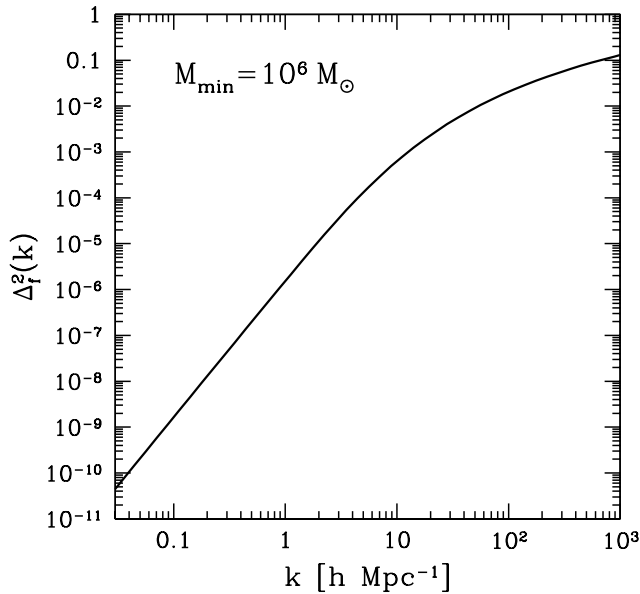}
\caption{Dimensionless power spectrum of $f=\delta^2 - \langle \delta^2
 \rangle$, $\Delta_{f,2}^2(k)$ [Eq.~(\ref{eq:dimpow})], from the
 two-point correlation term [Eq.~(\ref{eq:P_f,2})], evaluated at $z=0$ for
 $M_{\rm min} = 10^6 M_\odot$.}
\label{fig:Delta_2nd}
\end{center}
\end{figure}

The power spectrum, $P_f(k)$, is the Fourier transform of the two-point
correlation function of $f$ in real space, $\xi_f^{(2)}(\bm x - \bm y)
\equiv \langle f(\bm x) f(\bm y) \rangle$.
As $f$ is a quadratic function of $\delta$, $\xi_f^{(2)}$ is given by
the two- and four-point correlation functions of $\delta$, $\xi^{(2)}$
and $\xi^{(4)}$, as
\begin{eqnarray}
 \xi_f^{(2)}(\bm x - \bm y)&=&
  \xi^{(4)}(\bm x,\bm x,\bm y,\bm y)+2\xi^{(2)}(\bm x - \bm y)^2,
  \label{eq:two-point correlation function}
\end{eqnarray}
where we have suppressed the label $r$ (or equivalently $z$) to simply
the notation.
This motivates our decomposing the power spectrum into two parts as
\begin{equation}
 P_f(k) = P_{f,4}(k) + P_{f,2}(k),
\end{equation}
the first and second terms representing the contribution from the four-
and two-point correlation parts, respectively.
Since $\xi^{(2)}$ is related to the power spectrum of density
fluctuations, $P(k)$, through the Fourier transformation, the second
term, $P_{f,2}(k)$, is given by
\begin{eqnarray}
 P_{f,2}(k)
 &=& 2\int\frac{d^3p}{(2\pi)^3}\
  P(p)P(|\bm k-\bm p|).
  \label{eq:P_f,2}
\end{eqnarray}
Note that $P(k)$ is the {\it non-linear} power spectrum of density
fluctuations.
We calculate it using a halo model approach \cite{AnalyticHaloModel}
(see also Appendix~\ref{sec:Power spectrum of density fluctuation}).
Figure~\ref{fig:Delta_2nd} shows the two-point correlation contribution
to the ``dimensionless power spectrum'' of $f$, evaluated at $z=0$
for $M_{\rm min} = 10^6 M_\odot$; the quantity $\Delta_f^2(k)$ is
defined by
\begin{equation}
 \Delta_f^2(k) \equiv \frac{k^3}{2\pi^2} 
\frac{P_f (k)}{\langle \delta^2 \rangle^2}.
\label{eq:dimpow}
\end{equation}

\subsubsection{Four-point contribution}

\begin{figure}
\begin{center}
\includegraphics[width=8.5cm]{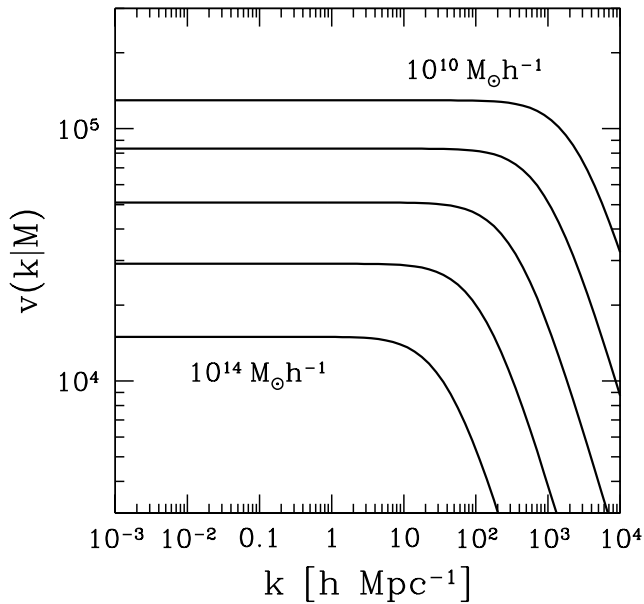}
\caption{The function that determines the four-point contribution to the
 power spectrum of CGB anisotropy, $v(k|M)$ [see Eqs.~(\ref{eq:PS 1-halo
 term}) and (\ref{eq:PS 2-halo term})], as a function of $k$ evaluated
  at $z=0$  for various values of $M$. The value of $M$ labeling each
 curve is spaced by one order of magnitude, with the largest and
 smallest values as indicated.}
 \label{fig:v}
\end{center}
\end{figure}

The four-point correlation function in Eq.~(\ref{eq:two-point
correlation function}) is further divided into 1-halo and 2-halo
contributions,
\begin{equation}
 \xi^{(4)}(\bm x,\bm x,\bm y,\bm y) = I_{1111}^{(1)} + I_{1122}^{(2)}.
\end{equation}
The ``1-halo term'' represents correlations between particles
within the same halo, whereas the ``2-halo term'' represents
correlations between particles in two distinct halos.
In other words, two points considered, ${\bm x}$ and ${\bm y}$, are
in one same halo for the 1-halo term, and are in two distinct halos
for the 2-halo term.\footnote{The full
expression of $\xi^{(4)}$ contains one 1-halo, six 2-halo, seven 3-halo,
and one 4-halo terms \cite{SB91}. However, most of them are vanishingly
small for a particular configuration considered here, $\xi^{4}(\bm x,\bm
x,\bm y,\bm y)$, as halos are spatially exclusive --- the same reason as
that stated in the second footnote.}
The corresponding power spectrum will be denoted as $P_{f,4}(k) =
P_{f,4}^{1h}(k) + P_{f,4}^{2h}(k)$.
According to Ref.~\cite{SB91}, the 1-halo and 2-halo terms of the
two-point correlation function are given respectively by
\begin{eqnarray}
 I_{1111}^{(1)} &=&
  \int_{M_{\rm min}}^\infty dM\ \frac{dn}{dM}\left(\frac{M}{\Omega_m
  \rho_c}\right)^4
  \int d\bm x_1
  \nonumber\\&&{}\times
  u^2(\bm x-\bm x_1|M) u^2(\bm y-\bm x_1|M),
  \label{eq:1-halo term}\\
 I_{1122}^{(2)} &=&
  \int_{M_{\rm min}}^\infty d M_1\ \frac{dn}{dM_1}
  \int_{M_{\rm min}}^\infty d M_2\ \frac{dn}{dM_2}
  \left(\frac{M_1M_2}{\Omega_m^2 \rho_c^2}\right)^2
  \nonumber\\&&{}\times
  \int d\bm x_1\int d\bm x_2\ u^2(\bm x-\bm x_1|m_1)
  \nonumber\\&&{}\times
  u^2(\bm y-\bm x_2|m_2) \xi_{hh}^{(2)}(\bm x_1,\bm x_2|M_1,M_2),
  \label{eq:2-halo term}
\end{eqnarray}
where $\xi_{hh}^{(2)}(\bm x_1,\bm x_2|M_1,M_2)$ is the two-point
correlation function of halos of mass $M_1$ and $M_2$, which is given by
$\xi_{hh}^{(2)}(r|M_1,M_2) \approx b(M_1) b(M_2) \xi_{\rm
lin}^{(2)}(r)$ [$b(M)$ is the linear halo bias and $\xi_{\rm lin}^{(2)}$
is the linear correlation function].
Equations~(\ref{eq:1-halo term}) and (\ref{eq:2-halo term}) are then
Fourier transformed, giving the power spectrum  as follows:
\begin{eqnarray}
 P_{f,4}^{1h}(k) &=&
  \int_{M_{\rm min}}^\infty dM\ \frac{dn}{dM}
  \left(\frac{M}{\Omega_m\rho_c}\right)^2 v^2(k|M),
  \label{eq:PS 1-halo term}\\
 P_{f,4}^{2h}(k) &=&
  \left[\int_{M_{\rm min}}^\infty dM\ \frac{dn}{dM}
   \left(\frac{M}{\Omega_m\rho_c}\right)
  b(M) v(k|M)\right]^2
  \nonumber\\&&{}\times
  P^{\rm lin}(k),
  \label{eq:PS 2-halo term}
\end{eqnarray}
where $v(k|M)$ is the Fourier transform of $u^2(\bm x|M) M / \Omega_m
\rho_c$, which is analytically solvable for the NFW profile with fairly
a lengthy expression, and shown in Fig.~\ref{fig:v} for various values
of $M$.
Note that these expressions are identical to those for the ordinary
matter power spectrum, Eqs.~(\ref{eq:1-halo PS}) and (\ref{eq:2-halo
PS}), if we replace $v(k|M)$ by $u(k|M)$.

\begin{figure}
\begin{center}
\includegraphics[width=8.5cm]{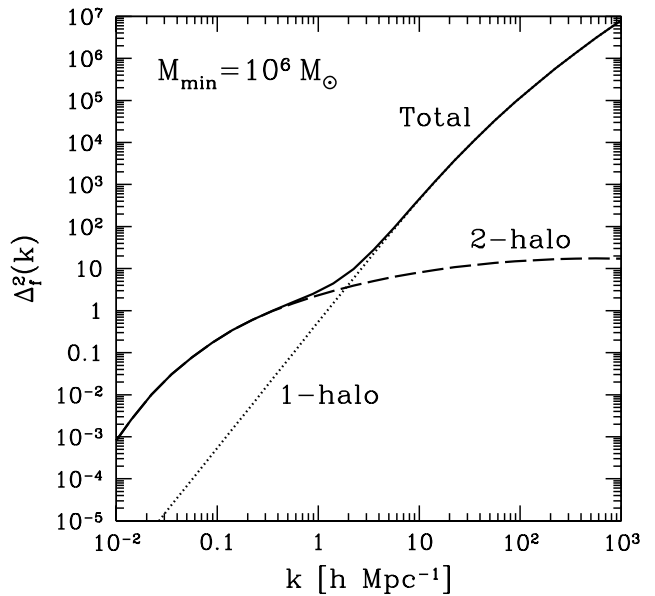}
\caption{Dimensionless power spectrum of
 $f=\delta^2-\langle\delta^2\rangle$ from the four-point contribution,
 $\Delta_{f,4}^2(k)$ [Eq.~(\ref{eq:dimpow})], by the 1-halo (dotted) and
 2-halo (dashed) terms [Eqs.~(\ref{eq:PS 1-halo term}) and (\ref{eq:PS
 2-halo term})]. Both are evaluated at $z=0$ for 
$M_{\rm min} = 10^6 M_\odot$.}
\label{fig:Delta}
\end{center}
\end{figure}

Figure~\ref{fig:Delta} shows the dimensionless power spectrum from the
four-point contribution, $\Delta^2_{f,4}(k)$, evaluated at $z=0$ for 
$M_{\rm min}= 10^6 M_\odot$.
We find that the four-point term always dominates over the two-point
term at all wavenumbers; thus, the two-point contribution may be safely
ignored. The 1-halo term dominates at smaller spatial scales,
$k\gtrsim 2~h~{\rm Mpc}^{-1}$, whereas the 2-halo term dominates
at larger scales.

\begin{figure}
\begin{center}
\includegraphics[width=8.5cm]{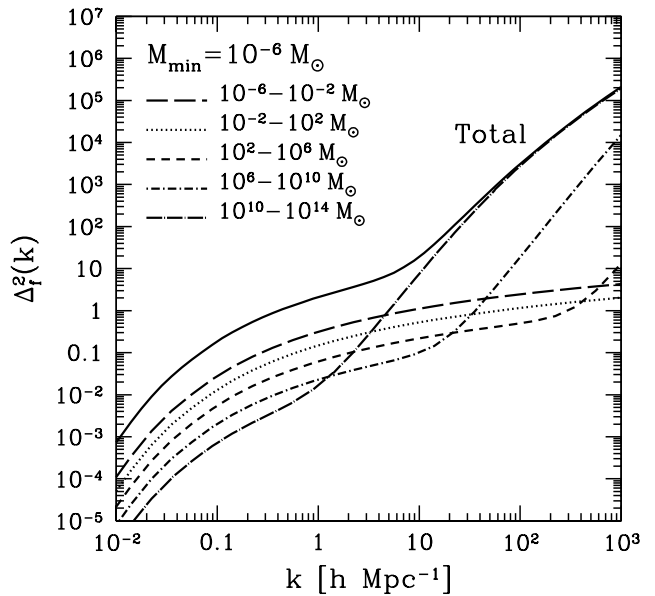}
\caption{The same as Fig.~\ref{fig:Delta} but for the smaller 
minimum mass, $M_{\rm min} = 10^{-6} M_\odot$. 
Contributions from halos of various mass ranges are
 also shown.}
\label{fig:Delta_mass}
\end{center}
\end{figure}

In Fig.~\ref{fig:Delta_mass}, we plot $\Delta_f^2(k)$ at $z = 0$ for
the smaller minimum mass, $M_{\rm min} = 10^{-6} M_\odot$.
The contributions from halos whose masses are in some specific ranges
are also shown.
We find an interesting trade-off between large and small mass halos.
Larger mass halos give much larger 1-halo term contributions and thus
completely dominate $\Delta^2_f(k)$ at small scales.
On the other hand, smaller mass halos give much larger 2-halo term
contributions and thus completely dominate $\Delta^2_f(k)$ at large scales.
This phenomena results in a ``break'' in the shape of  $\Delta^2_f(k)$
at a critical wavenumber, $k\sim 1$--$10~h~{\rm Mpc}^{-1}$ depending 
on $M_{\rm min}$, below which the dominant contribution to
 $\Delta^2_f(k)$ comes from the smallest halos, and above which
the dominant contribution comes from the largest halos.

We find by comparing Figs.~\ref{fig:Delta} and \ref{fig:Delta_mass} 
that the 2-halo contribution
to $\Delta^2_f(k)$
is virtually unaffected by $M_{\rm min}$. This is somewhat surprising,
given that the 2-halo contribution is dominated by the 
smallest halos having $M_{\rm min}$.
The reason for this phenomenon is because we define 
 $\Delta_f^2(k)$ as a ratio of $P_f(k)$ and the clumping factor squared,
$\langle\delta^2\rangle^2$  [see Eq.~(\ref{eq:dimpow})].
Reducing the minimum mass increases {\it both}  $P_f(k)$ and
$\langle\delta^2\rangle^2$ by the same amount, which makes 
the 2-halo contribution nearly independent of the choice of $M_{\rm min}$.
On the other hand, the 1-halo contribution decreases as $M_{\rm min}$
decreases. This is easy to understand. 
As the 1-halo contribution
to $P_f(k)$ is entirely dominated by the largest halos, 
it is independent of $M_{\rm min}$; however, as the clumping 
factor increases for smaller  $M_{\rm min}$,  the dimensionless power spectrum,
$\Delta_f^2(k)\propto P_f(k)/\langle\delta^2\rangle^2$, decreases
for smaller $M_{\rm min}$.
As a result, a smaller $M_{\rm min}$ makes the shape of $\Delta_f^2(k)$
flatter, reducing the power at larger $k$.
A ``shoulder'' due to the 2-halo term then becomes more
prominent for smaller $M_{\rm min}$, 
which is a characteristic that may be observable in the angular
anisotropy, as we discuss in the next section.

\section{Angular power spectrum}
\label{sec:Angular power spectrum}

By using Eq.~(\ref{eq:C_l}) with the 3D power spectrum shown in
Figs.~\ref{fig:Delta} and \ref{fig:Delta_mass}, we can calculate the
angular power spectrum, $C_l$, as a function of multipoles, $l$, for a
given dark matter model.
We here note that $C_l$ roughly corresponds to the correlation between
two points on the sky separated by an angle $\theta \approx \pi/l$.

\subsection{Supersymmetric neutralino}

First, we consider supersymmetric neutralinos as the dark matter
candidate, and assume that the neutralino mass is 100 GeV.
While the mean intensity, $\langle I_\gamma\rangle$, is very sensitive
to the dark matter mass, $\langle I_\gamma\rangle\propto m_\chi^{-2}$,
this factor exactly cancels out when the amplitude of anisotropy is
divided by the mean intensity.
Thus, $C_l$ depends on $m_\chi$ only weakly through the mass dependence
of the gamma-ray spectrum per annihilation.
We evaluate the anisotropy power per logarithmic range of $l$, $l (l+1)
C_l/2\pi$.

\begin{figure}
\begin{center}
\includegraphics[width=8.5cm]{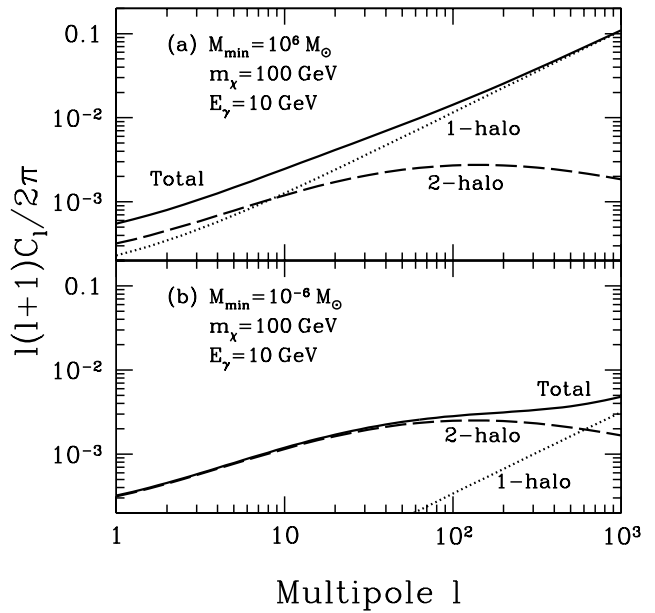}
\caption{
 Angular power spectrum of the CGB, $C_l$, from annihilation of
 supersymmetric neutralinos, evaluated for (a) $M_{\rm min} = 10^6
 M_\odot$ and (b) $M_{\rm min}=10^{-6} M_\odot$. Note that $C_l$ is
 dimensionless:  the mean intensity squared should be multiplied in
 order to convert it to the units of intensity squared. The neutralino
 mass $m_\chi$ is assumed to be 100 GeV. The predicted angular spectrum
 is shown at the observed gamma-ray energy of $E_\gamma = 10$
 GeV. Contributions to $C_l$ from the 1-halo (dotted) and 2-halo
 (dashed) terms are shown as well as the total signal (solid).}
\label{fig:C_l_halo}
\end{center}
\end{figure}

In Figs.~\ref{fig:C_l_halo}(a) and \ref{fig:C_l_halo}(b), we show the
predicted angular power spectrum evaluated at the observed gamma-ray
energy of $E_\gamma = 10$ GeV for $M_{\rm min} = 10^6 M_\odot$ and
$10^{-6} M_\odot$, respectively.
One can understand the shape of the angular spectrum for both 
cases using the same argument for understanding the shape of 
the 3D dimensionless power spectrum, $\Delta^2_f(k)$, in the previous
section. A rapid increase in $C_l$ at small angular scales (large $l$)
is due to the 1-halo term contribution. The 1-halo term dominates
even at relatively small $l$, $l\sim 10$, for the larger minimum mass,
whereas it is suppressed significantly for the smaller minimum mass,
in agreement with the dependence of the shape of  $\Delta^2_f(k)$
on $M_{\rm min}$ that we discussed in the previous section.
On the other hand, the 2-halo term contribution is nearly independent 
of $M_{\rm min}$, which is also in agreement with $\Delta^2_f(k)$.
(The 1-halo term is roughly proportional to the inverse of the
clumping factor squared, while the 2-halo term is nearly
independent of the clumping factor.)
This peculiar dependence of the shape of $C_l$ on $M_{\rm min}$
opens up an exciting possibility that one can ``measure'' the minimum
mass from the shape of $C_l$. Together with the information
from the energy spectrum of the CGB, therefore, it may be possible
to identify the mechanism by which gamma rays are produced
from annihilation of dark matter and the degree to which 
small halos are tidally disrupted by the structure formation.

Is the predicted angular power spectrum detectable? 
We compare the predicted power spectrum with the expected sensitivity
of the GLAST experiment.
We take the following specifications for GLAST: the field of view is
$\Omega_{\rm fov} = 4 \pi f_{\rm fov} = 2.4$ sr, the angular resolution
is $\sigma_b = 0.115^\circ$, and the effective area is
$A_{\rm eff} = 10^4$ cm$^2$ at 10 GeV \cite{GLAST}.
Note that the angular resolution is defined as a half width of a 
Gaussian point spread function, 
and the full width at half maximum is given by $\sqrt{8\ln 2}\sigma_b$.
In addition, for the diffuse gamma-ray observation, the background
contamination can be reduced to 6\% of the CGB, which is a promising
characteristic \cite{GLAST}.
Therefore, a fractional error of $C_l$,
\begin{equation}
 \frac{\delta C_l}{C_l}
  = \sqrt{\frac{2(1 + C_{N,\gamma} / W_l^2 C_l)^2}{(2l+1) \Delta l
  f_{\rm fov}}},
  \label{eq:C_l error}
\end{equation}
is essentially determined by the Poisson noise of the cosmic signal.
Here $\Delta l$ is the bin width of $l$ (we choose $\Delta l = 0.3 l$
for $M_{\rm min} = 10^6M_\odot$ and $0.5 l$ for $10^{-6}M_\odot$),
$C_{N,\gamma} = \Omega_{\rm fov} [(N_N/N_\gamma)^2 / N_N + 1/N_\gamma]$
is the power spectrum of photon noise, $N_N$ and $N_\gamma$ are the
count number of backgrounds and the signal ($N_N/N_\gamma \ll 1$ for
GLAST), respectively, and $W_l$ is the window function of a 
Gaussian point spread function,
$W_l = \exp(-l^2 \sigma_b^2 / 2)$.
If we assume that the CGB detected in GeV region is dominated by gamma
rays from dark matter annihilation, we may use the observed CGB
intensity, $E_\gamma^2\langle I_\gamma\rangle = 1.5 \times 10^{-6}$ GeV
cm$^{-2}$ s$^{-1}$ sr$^{-1}$ \cite{GeVCGB1,GeVCGB2}, to estimate the
signal-to-noise for anisotropy.
The expected number of photons, $N_\gamma$, for GLAST would then be
$N_\gamma = E_\gamma\langle I_\gamma\rangle \Omega_{\rm fov} A_{\rm eff}
t = 10^5 (t/1~\mathrm{yr})$ at $E_\gamma = 10$ GeV, while $N_N$ is
negligible.

\begin{figure}
\begin{center}
\includegraphics[width=8.5cm]{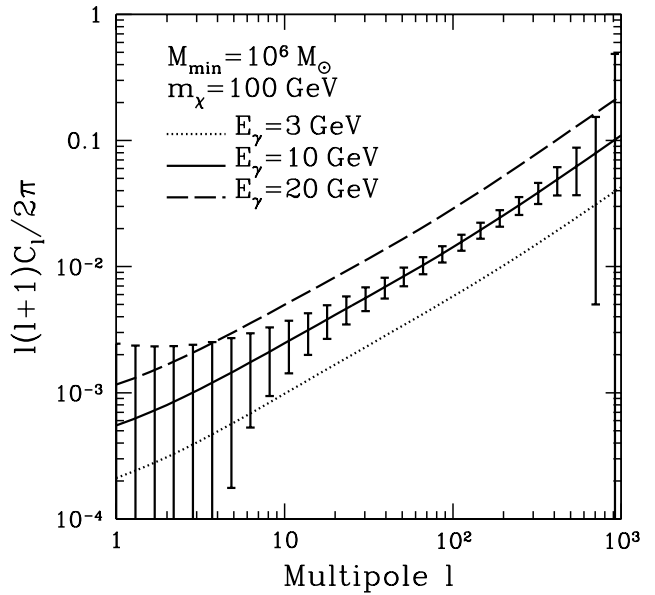}
\caption{Angular power spectrum of the CGB, $C_l$, from annihilation of
 supersymmetric neutralinos, evaluated for $M_{\rm min} =
 10^6M_\odot$. The neutralino mass $m_\chi$ is assumed to be 100
 GeV. The predicted angular spectrum is shown at the observed gamma-ray
 energy of $E_\gamma = 3$, 10, and 20 GeV. The $1\sigma$ error bars of
 $C_l$ expected from GLAST for 1 year of operation are also shown at
 $E_\gamma = 10$ GeV.}
\label{fig:C_l_cut6}
\end{center}
\end{figure}

\begin{figure}
\begin{center}
\includegraphics[width=8.5cm]{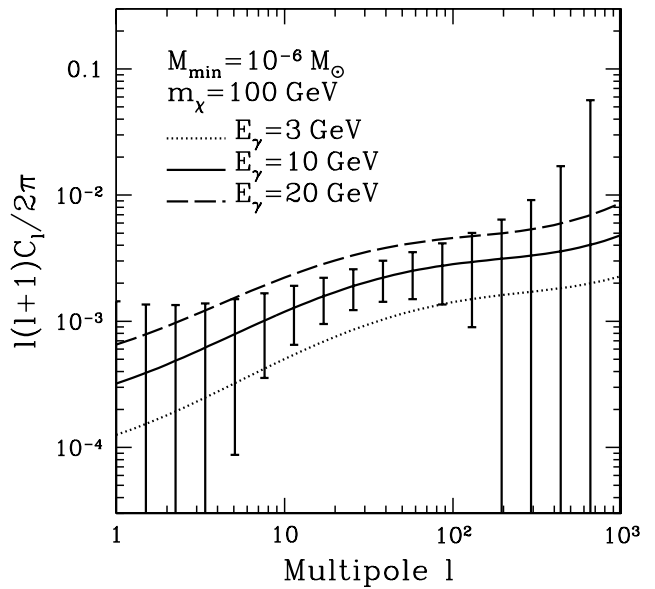}
\caption{The same as Fig.~\ref{fig:C_l_cut6}, but evaluated for the
 smaller
minimum mass, $M_{\rm  min} = 10^{-6}M_\odot$, which
is closer to the free-streaming mass of neutralinos.
(Note that the precise value of the free-streaming mass depends strongly
on the dark matter mass and interaction properties.)}
\label{fig:C_l_cut-6}
\end{center}
\end{figure}

In Fig.~\ref{fig:C_l_cut6}, we show the predicted angular power spectrum
at the observed gamma-ray energies of $E_\gamma=3$, 10, and 20 GeV, 
assuming $M_{\rm min}=10^6M_\odot$, with 
the expected $1\sigma$ errors of
$C_l$ at $E_\gamma = 10$ GeV for $t = 1$ yr of observations.
We find that the GLAST should be able to measure the angular power
spectrum of the CGB fairly easily for 1 year of observations, if the
dark matter particle is the neutralino with mass around 100 GeV and its
annihilation dominates the observed CGB in GeV region.
The angular power spectrum for the smaller minimum mass, 
$M_{\rm min} = 10^{-6}M_\odot$
is shown in 
Fig.~\ref{fig:C_l_cut-6}, with the expected error
bars.
We find that anisotropy is still easily detectable with the GLAST for 
1 year of observations.
Therefore, we conclude that, if dark matter particles are 
supersymmetric neutralinos and the observed CGB in GeV region
is dominated by their annihilation, 
the GLAST should be able to measure the angular power spectrum
of CGB anisotropy, regardless of the minimum mass.

As prospects for detecting CGB anisotropy in the GLAST data 
are very good, it is tempting to speculate that one
may actually detect CGB anisotropy in the existing EGRET data.
The EGRET parameters are $\Omega_{\rm fov} = 0.5$ sr, $\sigma_b =
0.5^\circ$, and $A_{\rm eff} = 7 \times 10^2$ cm$^2$ at 10 GeV
and on-axis direction
($A_{\rm eff}$ decreases by a factor of 2 at 20$^\circ$ and
of 6 at 30$^\circ$) \cite{EGRET}.
The background noise has been suppressed by an order of magnitude compared
with the photon counts from the CGB.
Although the detector sensitivity of the EGRET is not as good as 
that of the GLAST, 
we expect that the Poisson error (photon noise) of the
signal to be quite small at $l \alt 400$ owing to 
a longer integration time (EGRET has observed
the sky for much more than 1 year).
The data above $l \approx 400$ would not be very useful 
because of the limited angular resolution.
While one needs to carefully characterize the systematic uncertainty 
due to the Galactic foreground removal, 
the power spectrum of CGB anisotropy in the EGRET
data should be measured in search of the signatures of dark
matter annihilation.

\subsection{MeV dark matter}

\begin{figure}
\begin{center}
\includegraphics[width=8.5cm]{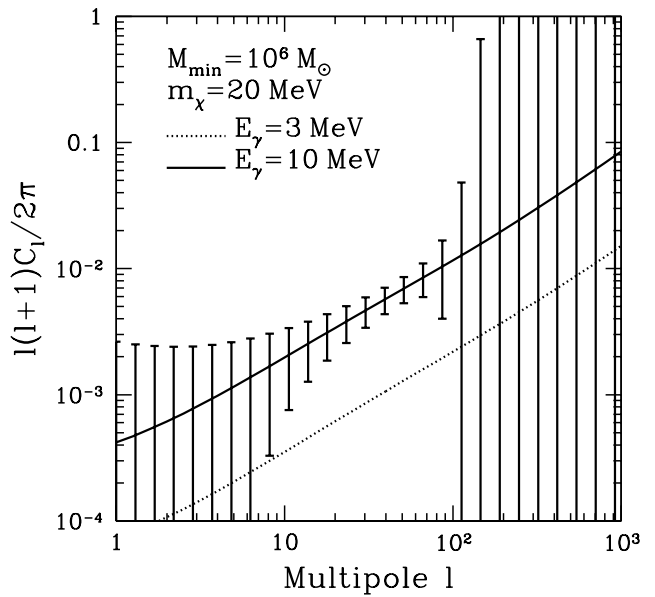}
\caption{The same as Fig.~\ref{fig:C_l_cut6} but for MeV dark matter with
 $m_\chi = 20$ MeV, evaluated at $E_\gamma = 3$ (dotted) and 10 (solid)
 MeV. Error bars are calculated for ACT in 1 year of operation.}
\label{fig:C_l_MeV_cut6}
\end{center}
\end{figure}

As for the MeV dark matter, the continuum gamma rays are emitted by the
internal bremsstrahlung, and the spectrum shape is shown in
Sec.~\ref{sec:Cosmic gamma-ray background: Isotropic component} (see
also, Refs.~\cite{AK2,BBB}).
Figure~\ref{fig:C_l_MeV_cut6} shows the predicted 
angular power spectrum of CGB anisotropy for the MeV dark matter
with $m_\chi = 20$ MeV. We show $C_l$ at the observed gamma-ray energy
of $E_\gamma = 3$ and 10 MeV, as well as the expected uncertainty
at 10 MeV.
The shape as well as the normalization are very similar to those for
the neutralinos.
In the MeV regime, the most promising technique to detect gamma rays is
the Compton scattering inside the detector.
Although the design of future detectors of this kind, such as the 
Advanced Compton Telescope (ACT) \cite{ACT}, are still in progress, we 
parameterize it with $\sigma_b = 1^\circ$, $A_{\rm eff} = 100$ cm$^2$,
$\Omega_{\rm fov} = 1$ sr, $t = 1$ yr, and $N_N/N_\gamma = 1$
\cite{SNIaAnisotropy}.
Again taking the mean CGB intensity at 10 MeV to be what has been
observed, $10^{-4}$ cm$^{-2}$ s$^{-1}$ sr$^{-1}$
\cite{MeVCGB1,MeVCGB2,MeVCGB3}, the signal count with ACT would be
$N_\gamma = 3 \times 10^5 (t/1\ \mathrm{yr})$.
With these parameters for ACT, we find that it should be easy to measure
the power spectrum in the range of $10 < l < 100$.
Above $l = 100$, the errors become exponentially larger because of the
detector angular resolution, i.e., due to the effect of the window
function $W_l$ in Eq.~(\ref{eq:C_l error}); this is unavoidable unless
better angular resolution is realized.

\begin{figure}
\begin{center}
\includegraphics[width=8.5cm]{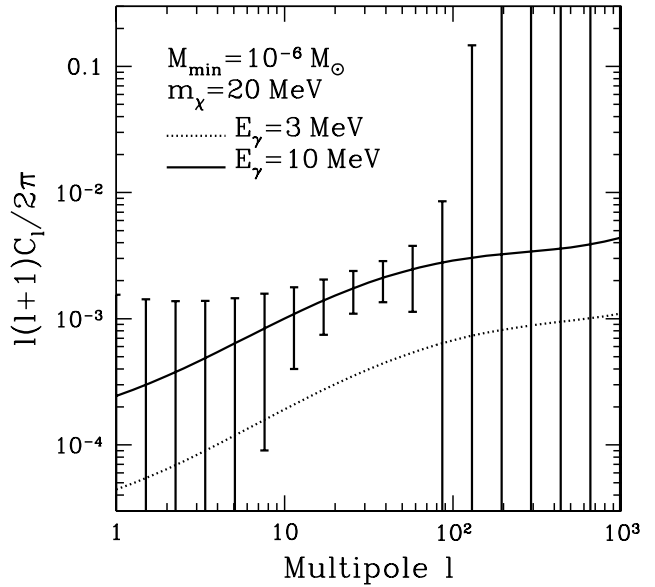}
\caption{The same as Fig.~\ref{fig:C_l_MeV_cut6}, but evaluated for
the smaller minimum mass, $M_{\rm min} = 10^{-6}M_\odot$, which
is closer to the free-streaming mass of MeV dark matter.
(Note that the precise value of the free-streaming mass depends strongly
on the dark matter mass and interaction properties.)}
\label{fig:C_l_MeV_cut-6}
\end{center}
\end{figure}

The dependence of $C_l$ 
on the minimum mass, $M_{\rm min}$, is almost the same as that for
the neutralino.
In Fig.~\ref{fig:C_l_MeV_cut-6},
we show the angular power spectrum for the smaller minimum mass, 
$M_{\rm min} = 10^{-6}M_\odot$.
(And this is a correct mass to use, up to the dependence on the dark
matter mass \cite{AK1}, when the emission is due to
the internal bremsstrahlung; for the minimum mass appropriate
for the 511 keV emission, see arguments in Ref.~\cite{rasera/etal:2005}.)
Again, the predicted angular power spectrum is well above the expected 
uncertainty, and thus our conclusion about prospects for detection
of anisotropy is robust regardless of the minimum mass.

The COMPTEL has measured the CGB in this energy region \cite{MeVCGB3}.
The parameters of the COMPTEL are $\Omega_{\rm fov} = 1$ sr, $\sigma_b =
1.5^\circ$, $A_{\rm eff} = 30$ cm$^2$ at 20 MeV \cite{COMPTEL}, 
which are almost comparable to the parameters adopted for the ACT.
The biggest issue is, however, its poor signal-to-background ratio,
$N_\gamma / N_N \approx 0.01$; thus, it would be quite challenging to
detect CGB anisotropy in the COMPTEL data because of
large statistical errors.

\section{Discussion}
\label{sec:Discussion}

\subsection{Dependence on gamma-ray energy}
\label{sub:Dependence on gamma-ray energy}

\begin{figure}
\begin{center}
\includegraphics[width=8.5cm]{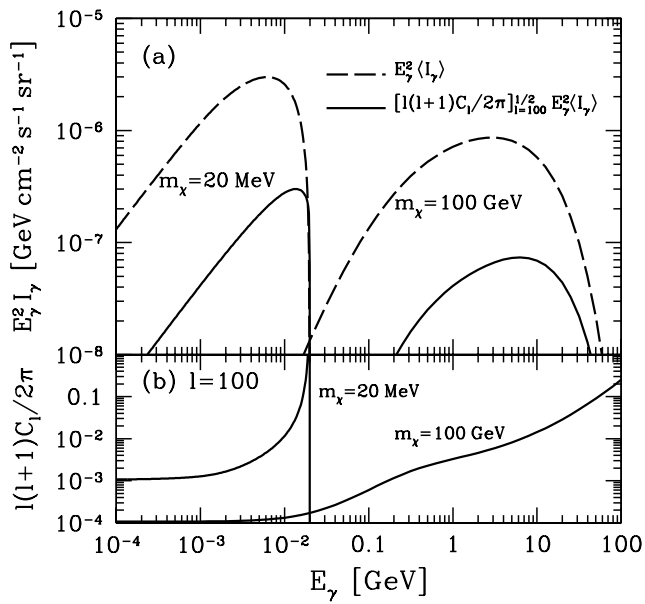}
\caption{(a) Energy spectrum of isotropic (dashed) and anisotropic 
 (solid) component of the
 CGB intensity for $m_\chi = 100$ GeV (right curves) and 20 MeV 
 (left curves), evaluated for $M_{\rm min} = 10^6 M_\odot$. 
 Anisotropic components are evaluated at
 $l = 100$. 
 (b) Energy spectrum of anisotropy divided by the mean intensity
 at $l = 100$.}
\label{fig:anis_spect}
\end{center}
\end{figure}

In addition to the shape of the angular power spectrum as a function of
$l$, it would be useful to discuss dependence on the gamma-ray energy in
order to obtain further information on the CGB origin.
We show the energy spectrum of isotropic, $\langle I_\gamma\rangle$,
and anisotropic, $\langle I_\gamma\rangle [l(l+1)C_l/(2\pi)]^{1/2}$, 
components of the CGB in units of intensity 
in Fig.~\ref{fig:anis_spect}(a),
for both the neutralino
($m_\chi = 100$ GeV) and MeV dark matter ($m_\chi = 20$ MeV) assuming
$M_{\rm min} = 10^6 M_\odot$.
The energy spectrum of anisotropy divided by the mean intensity,
$l(l+1)C_l/(2\pi)$, is also shown in 
Fig.~\ref{fig:anis_spect}(b).
We show only $l=100$.
The isotropic spectrum is boosted accordingly in order to explain the
observed intensity; the details are given in the last paragraph
of Sec.~II.

We find that anisotropy is enhanced when the spectrum has sharp 
features such as edges (as seen at
$\sim 20$ MeV for the MeV dark matter case) or lines. 
This has been pointed out by Ref.~\cite{SNIaAnisotropy} in the context of
CGB anisotropy from Type Ia supernovae.
Therefore, if the energy resolution of gamma-ray 
detectors is sufficiently good to resolve these properties, 
the spectral features would provide another powerful diagnosis
of dark matter annihilation signals.
The features might also allow us to obtain information on
the redshift evolution of the sources (see, Ref.~\cite{SNIaAnisotropy}
for more details).
Other potential spectrum signatures include the line gamma emission
($\chi\chi \to \gamma\gamma, Z^0\gamma$) and internal bremsstrahlung
($\chi\chi \to W^- W^+ \gamma$ \cite{NeutralinoBrems}) from annihilation
of neutralinos, as well as the 511 keV line emission from
annihilation of MeV dark matter particles into $e^-e^+$ pairs.

\subsection{Other astrophysical sources}
\label{sub:Other astrophysical sources}

The angular power spectrum shown in
Figs.~\ref{fig:C_l_cut6}--\ref{fig:C_l_MeV_cut-6} should be very
characteristic of annihilating dark matter in extragalactic dark matter
halos, as the gamma-ray intensity is proportional to the density
squared.
The intensity of gamma rays coming directly from other astrophysical
sources should be linearly proportional to density.
It is likely that blazars are the most dominant constituent of the GeV
gamma rays over a wide energy range \cite{Blazar1,Blazar2}.
Assuming that blazars are biased tracers of the underlying mass
distribution, the two-point correlation function of blazars should be
simply given by that of density fluctuations, $P(k)$.
For more quantitative study, we perform the following
simple analyses for the power spectrum of blazars.

\begin{figure}
\begin{center}
\includegraphics[width=8.5cm]{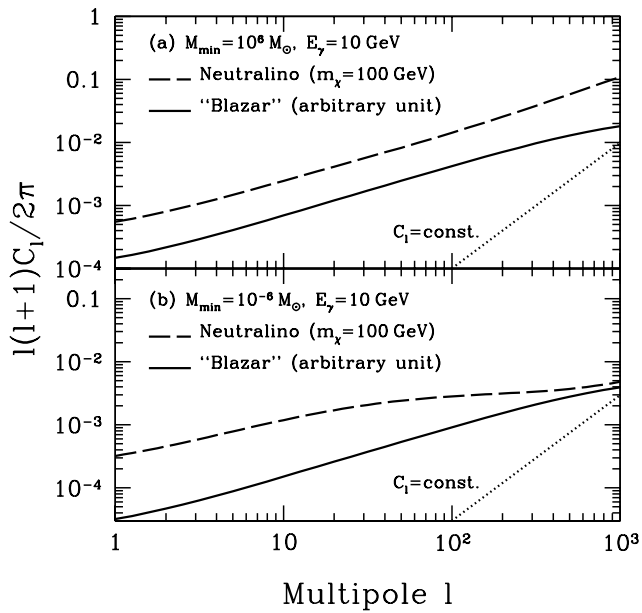}
\caption{Shape of the angular power spectrum of the CGB expected 
 from unresolved ``blazars''
 (solid lines; 
 see the footnote 5 for the reason why the quotation marks are put)
 with arbitrary normalizations.
 The power spectrum from annihilation of neutralinos 
 with $m_\chi = 100$ GeV is also plotted as the dashed lines. 
 The adopted gamma-ray energy is
 10 GeV, and the minimum mass of dark matter halo 
 is (a) $10^6 M_\odot$, and (b) $10^{-6} M_\odot$. 
 The dotted lines show the shot 
 noise ($C_l={\rm const.}$) with arbitrary  normalizations,
 which represent the power spectrum of very rare sources.}
\label{fig:C_l_blazar}
\end{center}
\end{figure}

First, we assume that blazars are very rare objects and 
their angular spectrum is entirely dominated by the shot noise.
In this case the angular power spectrum does not depend on $l$,
and thus $l(l+1)C_l$ is proportional to $l^2$ at $l\gg 1$. 
We show in Figs.~\ref{fig:C_l_blazar}(a) and \ref{fig:C_l_blazar}(b) 
that the shot noise spectrum totally lacks the power on large
angular scales (i.e., the spectrum is too steep), and can be 
easily distinguished from dark matter annihilation.

Second, we take the other extreme limit where blazars are 
quite common and trace the underlying matter density field, $\delta$,
fairly well.
Specifically, 
we calculate 
 the line-of-sight integral in
Eq.~(\ref{eq:intensity}) with  $\delta^2$ replaced by $\delta$.
The volume emissivity is then given by 
$P_\gamma \propto \delta (1+z)^3 E_\gamma
dN_\gamma / dE_\gamma$, and we assume a power-law
blazar energy spectrum falling off as  
$dN_\gamma / dE_\gamma \propto E_\gamma^{-2}$ (e.g.,
\cite{Mrk421} for Mrk 421),
which gives $W(E_\gamma,z) \propto E_\gamma^{-2} e^{-\tau(E_\gamma,
z)}$.
We calculate the angular power spectrum, $C_l$, using
Eq.~(\ref{eq:C_l}) with $P_f(k)$ replaced by $P(k)$.
Finally, we assume that blazars form only when the mass of 
host dark matter halos is larger than $M_{\rm min}=10^{11} M_\odot$.
In this simplified prescription, the average number of blazars
in a halo linearly increases with the host halo mass.
In reality, however, this may not be true and we may also need
to take into account the difference between a central galaxy and
satellite galaxies within a dark matter halo.
This kind of galaxy distribution model, called the Halo Occupation
Distribution (see Ref.~\cite{HaloModelReview} for a review), 
could be made more realistic for blazars; however,
we do not pursue it in any more detail, and adopt the above model
for calculating the angular power spectrum of gamma-ray emission from
``blazars.''\footnote{Our model is probably not a very good model for 
describing the power spectrum of real blazars. It simply describes 
the angular power spectrum of dark matter halos above 
$10^{11} M_\odot$ which emit gamma rays (by whatever mechanism) with 
an energy spectrum falling off as 
$dN_\gamma / dE_\gamma \propto E_\gamma^{-2}$.
In reality, not all dark matter halos above $10^{11} M_\odot$
host blazars, and blazars are known to be transient objects
which turn on and off.}

We show the shape of the angular power spectrum expected from blazars in
Figs.~\ref{fig:C_l_blazar}(a) and \ref{fig:C_l_blazar}(b).
(Normalizations are taken arbitrarily.)
The blazar power spectrum is determined by the 1-halo term
at all multipoles; thus, the spectrum is nearly a power law 
with a little bit
of flattening above $l\sim 200$, which is due to the minimum mass
of halos hosting blazars, $10^{11} M_\odot$. 
The small-scale power drops as we 
raise the minimum mass of host halos 
(see, e.g., Fig.~2 of Ref.~\cite{AnalyticHaloModel}).
In Fig.~\ref{fig:C_l_blazar}(a), we compare the blazar spectrum
with the spectrum of CGB anisotropy from annihilation of
neutralinos with $m_\chi = 100$ GeV for the large minimum mass,
 $10^6 M_\odot$. As we have noted before, the annihilation spectrum
for $M_{\rm min}=10^6 M_\odot$
is also dominated by the 1-halo term at all multipoles,
which makes the annihilation and blazar spectrum actually look similar
at $l\lesssim 200$. Above $l\sim 200$, however, the annihilation
spectrum continues to grow whereas the blazar spectrum flattens out.
In Fig.~\ref{fig:C_l_blazar}(b), we use the smaller minimum
mass, $10^{-6} M_\odot$, which is probably more realistic
for dark matter annihilation. In this case the dominant contribution
comes from the 2-halo term, which makes the annihilation spectrum
substantially flatter than the blazar spectrum.
In other words, the annihilation spectrum has much more power
at large angular scales, which should be easily distinguished
from the blazar spectrum.
The same argument should also apply to the CGB in the MeV region
from Type Ia supernovae. 
(See Ref.~\cite{SNIaAnisotropy} for another approach to 
calculating the angular correlation function of Type Ia supernovae.)

Another potential astrophysical contributor to the CGB
is the gamma-ray emission from particles accelerated by shock waves 
forming in galaxy clusters
\cite{ClusterShock1,ClusterShock1.5,ClusterShock2,ClusterShock3,ClusterShock4,ClusterShock5}.
There have been a few claims about detection of a correlation
between the position of clusters and the EGRET 
data \cite{ClusterGamma1,ClusterGamma2}.
When the gamma rays are emitted mainly by the inverse Compton
scattering off the CMB photons by accelerated electrons, which many
authors considered to be the dominant emission process, the
gamma-ray intensity would be proportional to the baryon density at the shock. 
On the other hand, when the gamma rays are produced by collisions between
cosmic-ray protons (accelerated by the cluster shocks) and intracluster
medium, the gamma-ray intensity would be proportional to 
the baryon density squared at the shock.
However, the baryon density distribution at the shock would be very
different from the distribution of dark matter particles.
One would need cosmological hydrodynamical simulations to study 
the shape of the angular power spectrum of CGB from such processes.

\subsection{Substructure of dark matter halos}
\label{sub:Substructure of dark matter halos}

The biggest uncertainty in calculations of the CGB, whether the mean
intensity or anisotropy, is the substructure (i.e., small-scale
clumpiness) within dark matter halos.
$N$-body simulations of collisionless dark matter universally indicate
the existence of sub-halo clumping of dark matter
\cite{DMSubstructure}.
This effect has been entirely ignored in our calculations, as we have
assumed that each halo has a smooth density profile.
It is expected that the dark matter substructure increases the gamma-ray
intensity by simply increasing the clumpiness.\footnote{For gamma rays
from annihilation of dark matter in the substructure of our Galaxy, see
Refs.~\cite{GalacticHalo1,GalacticHalo2,GalacticHalo3,GalacticHalo3.5,GalacticHalo4,GalacticHalo5,GalacticHalo6}.}
How much it is increased depends very much on the exact amount of
substructure, which is difficult to predict due to a limited resolution
of the current $N$-body simulations.
In this regard, our calculations provide the lower limit to the angular
power spectrum of CGB.
The fact that our predictions are already well above the noise level of
GLAST and ACT is extremely encouraging for prospects of detection of
anisotropy.

Let us comment on some effects of dark matter substructures on the
angular power spectrum.
It is expected that the effect of substructure would be smaller for
anisotropy than for the mean intensity, when the amplitude of anisotropy
is divided by the mean intensity (which is what we have been using to
calculate $C_l$).
But still, the increased clumping will increase the four-point function
of density fluctuations, particularly 1-halo term which dominates at
small scales.
Therefore, we expect the substructure to modify $C_l$ at large $l$,
while we do not expect a significant change in $C_l$ at small $l$, where
the four-point function is dominated by the 2-halo term.
Finally, in the limit that dark matter annihilation signals are 
completely dominated by the sub-halos within a bigger (host) halo,
e.g., Earth-mass micro halos in a Milky-Way size halo, 
the gamma-ray intensity may be {\it linearly} proportional to
the density profile of host halos \cite{OTN05}.
In this model, there is no gamma-ray emission from 
the intra-halo medium between the sub-halos,
and the sub-halos are assumed to follow the density profile
of a host halo. Our formalism should be modified for this case
by including the number of sub-halos within a host halo as a function
of the host halo mass, which is essentially the Halo Occupation
Distribution for the sub-halos.

Incorporating dark matter substructures into the halo model is an active
field of research, and therefore we expect to be able to get a better
handle on this effect in near future.

\section{Conclusions}
\label{sec:Conclusions}

The CGB provides indirect means to probe the nature of dark matter
particles via high-energy photons from dark matter annihilation.
The dark matter annihilation has occurred in all the past halos, and now
contributes to the CGB flux at some level; its contribution might be
dominant if the flux is boosted by some mechanism such as tidally
survived dark matter substructure.
Therefore, revealing the origin of the CGB is an very important problem
that is potentially connected to the dark matter properties and halo
substructure as well as ordinary astrophysical objects.
In order to achieve this purpose, while the energy spectrum of the mean intensity of CGB has been
investigated by many researchers, anisotropy of CGB  from dark matter
annihilation has been entirely neglected.
In this paper we have calculated the angular power spectrum of CGB from
dark matter annihilation for the first time, using an analytical halo
model approach.
As for dark matter candidates, we have discussed two possibilities: one
is the supersymmetric neutralino, one of the most popular candidates
today, which contributes to the CGB in GeV region.
The other is the MeV dark matter, a dark matter species first introduced
to explain the 511 keV emission line from the Galactic center by
annihilation into $e^-e^+$ pair, which contributes to the CGB in MeV
region.

Since the gamma-ray intensity from annihilation is proportional to the
density squared, $\rho_\chi^2$, we have derived the 3D power spectrum of
the quantity $f = \delta^2 - \langle \delta^2 \rangle$, $P_f(k)$.
The calculation involves the Fourier transformation of the four-point
correlation function of underlying mass density fluctuations,
$\xi^{(4)}$, which has been shown to dominate over the two-point
correlation contribution, $\xi^{(2)2}$.
This $P_{f,4} (k)$ includes the 1-halo and 2-halo terms, the former
containing two points within the same halo, and the latter containing
two points in two different halos.
The analytical expressions for $P_f(k)$ from each term are given in
Eqs.~(\ref{eq:P_f,2}), (\ref{eq:PS 1-halo term}), and (\ref{eq:PS 2-halo
term}), and the results are plotted in Figs.~\ref{fig:Delta_2nd},
\ref{fig:Delta}, and \ref{fig:Delta_mass}.
At all scales the four-point contribution totally dominates the signal; 
at small scales, the 1-halo term dominates.

We note that our formalism can also be used for any other emission
processes that involve collisions of two particles.
For example, one may use this to compute the power spectrum of free-free
or bound-free emission from the ionized gas in halos.
For this application one needs to replace the dark matter density
profile with the gas density profile.

Using Eq.~(\ref{eq:C_l}) that connects the 3D power spectrum, $P_f(k)$,
to the angular power spectrum, $C_l$, we have calculated the CGB angular
power spectrum as a function of multipoles, $l$, for both the neutralino
($m_\chi = 100$ GeV) and MeV dark matter ($m_\chi = 20$ MeV) at various
gamma-ray energies.
We have also compared the predicted signals with the expected
sensitivity of future gamma-ray detectors --- GLAST in GeV region (for
neutralinos) and ACT in MeV region (for MeV dark matter).
The results are shown in
Figs.~\ref{fig:C_l_cut6}--\ref{fig:C_l_MeV_cut-6}.
For both cases, we have found that these detectors will have sufficient
sensitivity to measure the angular power spectrum with reasonable
accuracy.

We have studied the effects of the minimum mass, $M_{\rm min}$, on the
predicted angular power spectrum in detail. 
While the 1-halo contribution, which dominates at small angular
scales (large $l$), 
decreases for smaller $M_{\rm min}$, the 2-halo contribution, which
dominates at large angular scales (small $l$), is virtually unaffected
by $M_{\rm min}$.
This property results in a peculiar dependence of the shape of $C_l$
on $M_{\rm min}$, which may be used in combination with the information
from the energy spectrum of the CGB to determine $M_{\rm min}$.
As $M_{\rm min}$ depends on the radiation processes of gamma rays from
annihilation as well as the survival of micro halos contributing 
to the CGB, the shape of $C_l$ provides a powerful tool for 
determining these properties, which are otherwise difficult to probe.
Our conclusion about prospects for measuring $C_l$ of CGB anisotropy are robust
regardless of $M_{\rm min}$.

By applying our formalism with some modification, we have shown that the
other astrophysical sources such as blazars would reveal a 
different shape of the angular power spectrum, as these contributions
are linearly proportional to density fluctuations.
The shape of $C_l$ at large $l$ might further change when we take into
account the dark matter substructure, but the result in this paper sets
a lower bound on anisotropy at all multipoles, which provides excellent
prospects for detection of CGB anisotropy by future gamma-ray
detectors.

We conclude that the angular power spectrum of CGB provides a
smoking-gun signature for gamma-ray emission from annihilation of dark
matter particles, which would be a powerful tool for understanding the
nature of dark matter particles.


\acknowledgments

We would like to thank Kyungjin Ahn, John Beacom, Gianfranco Bertone, 
Celine B{\oe}hm, Peter Michelson, and Tomonori Totani for valuable
comments. We would also like to thank Duane E. Gruber for providing
us with the COMPTEL data plotted in Fig.~\ref{fig:CGB_spect}.
S.~A. was supported by a Grant-in-Aid from the JSPS.
E.~K. acknowledges support from an Alfred P. Sloan Fellowship.

\appendix

\section{Relation between angular and three-dimensional power spectrum}
\label{eq:Relation between angular and three-dimensional power spectrum}

In this section, we derive the relation between the angular power
spectrum, $C_l$, and the 3D power spectrum, $P_f(k)$, that
we have given in Eq.~(\ref{eq:C_l}).
With the spatial Fourier transformation, Eq.~(\ref{eq:a_lm}) is
rewritten as
\begin{eqnarray}
 \langle I_\gamma \rangle a_{lm} &=& \int d \hat{\bm n} \int dr
  \int \frac{d\bm k}{(2\pi)^3}\
  \tilde f_{\bm k}(r) e^{i\bm k\cdot \bm r}
  W(r)Y_{lm}^\ast (\hat{\bm n})
  \nonumber\\
 &=& \int d\hat{\bm n} \int dr \int \frac{d\bm k}{2\pi^2}\
  \tilde f_{\bm k}(r)W(r) Y_{lm}^\ast (\hat{\bm n})
  \nonumber\\&&{}\times
  \sum_{l^\prime m^\prime} i^{l^\prime} j_{l^\prime} (kr)
  Y_{l^\prime m^\prime}^\ast (\hat{\bm k})
  Y_{l^\prime m^\prime} (\hat{\bm n})
  \nonumber\\
 &=& i^l \int dr\ W(r) \int \frac{d\bm k}{2\pi^2}\
  \tilde f_{\bm k}(r) j_l(kr) Y_{lm}^\ast (\hat{\bm k}),
  \nonumber\\
  \label{eq:a_lm 2}
\end{eqnarray}
where in the second equality $e^{i\bm k\cdot \bm r}$ is expanded into
the spherical harmonics and the spherical Bessel function, $j_l(kr)$, 
using Rayleigh's formula, and the last equality is due to the
orthonormal relation of $Y_{lm}(\hat{\bm n})$.
Hence, the angular power spectrum $C_l = \langle |a_{lm}|^2 \rangle$ is
expressed by
\begin{eqnarray}
 \langle I_\gamma \rangle^2 C_l
 &=&  \int dr\ W(r) \int dr^\prime\ W(r^\prime)
  \frac{2}{\pi} \int d\bm k\  P_f(k;r,r^\prime)
  \nonumber\\&&{}\times
  j_l(kr) j_l(kr^\prime)
  Y_{lm}(\hat{\bm k}) Y_{l^\prime m^\prime}^\ast (\hat{\bm k})
  \nonumber\\
 &=&  \int dr\ W(r) \int dr^\prime\ W(r^\prime)
  \nonumber\\&&{}\times
  \left[\frac{2}{\pi} \int k^2 dk\  P_f(k;r,r^\prime)
  j_l(kr) j_l(kr^\prime)\right],
  \nonumber\\
  \label{eq:C_l 1}
\end{eqnarray}
where we used the definition of $P_f$, $\langle \tilde f_{\bm k}(r)
\tilde f_{\bm k^\prime}^\ast (r^\prime) \rangle = (2\pi)^3 \delta^{(3)}
(\bm k - \bm k^\prime) P_f(k;r,r^\prime)$.
As a final step, using the following approximation,
\begin{eqnarray}
 \lefteqn{\frac{2}{\pi}\int k^2 dk\ P_f(k;r,r^\prime)
  j_l(kr)j_l(kr^\prime)}
  \nonumber\\
 &\simeq& \frac{1}{r^2}
  P_f\left(k = \frac{l}{r};r\right)
  \delta^{(1)} (r-r^\prime),
\end{eqnarray}
which is valid if $P_f(k;r,r^\prime)$ varies relatively slowly as a
function of $k$, we arrive at Eq.~(\ref{eq:C_l}).

\section{Power spectrum of density fluctuation}
\label{sec:Power spectrum of density fluctuation}

In this section, we summarize the power spectrum $P(k)$ of density
fluctuations, $\delta$.
The detailed derivation has been given in, e.g.,
Refs.~\cite{SB91,HaloModelReview,AnalyticHaloModel}, 
to which we refer the reader.
$P(k)$ includes two distinctive terms, $P(k) = P^{1h}(k) +
P^{2h}(k)$, each term given by
\begin{eqnarray}
 P^{1h}(k) &=& \int_{M_{\rm min}}^\infty dM\ \frac{dn}{dM}
  \left(\frac{M}{\Omega_m \rho_c}\right)^2 |u(k|M)|^2,
  \label{eq:1-halo PS}\\
 P^{2h}(k) &=& \left[\int_{M_{\rm min}}^\infty dM\ \frac{dn}{dM}
  \left(\frac{M}{\Omega_m \rho_c}\right) b(M) u(k|M)\right]^2
  \nonumber\\&&{}\times
  P_{\rm lin}(k),
  \label{eq:2-halo PS}
\end{eqnarray}
representing the 1-halo and 2-halo contributions, respectively.
Here $u(k|M)$ is the Fourier transform of $u(r|M)$, the analytic
representation for the NFW profile given by Eq.~(81) of
Ref.~\cite{HaloModelReview}, $b(M)$ is the bias parameter
\cite{HaloModelReview}, and $P_{\rm lin}$ is the linear power spectrum,
for which we adopt the fitting formula in Ref.~\cite{LinearPS}.
The values for the 1-halo and 2-halo terms as well as the linear power
spectrum are shown in, e.g., Fig.~1 of Ref.~\cite{AnalyticHaloModel}.
The 1-halo term dominates at small spatial scales (large $k$).

\end{document}